\documentclass[twocolumn,showpacs,prb,aps, superscriptaddress]{revtex4}
\usepackage{amssymb}
\usepackage{amsmath}
\usepackage{graphicx}

\begin{document}

\title{Current noise cross correlation mediated by Majorana bound states}

\author{Hai-Feng L\"{u}}

\affiliation{Department of Applied Physics, University of Electronic
Science and Technology of China, Chengdu 610054, China}

\affiliation{Department of Physics, The University of Hong Kong,
Pokfulam Road, Hong Kong, China}

\author{Hai-Zhou Lu}

\affiliation{Department of Physics, The University of Hong Kong,
Pokfulam Road, Hong Kong, China}

\author{Shun-Qing Shen}

\affiliation{Department of Physics, The University of Hong Kong,
Pokfulam Road, Hong Kong, China}

\begin{abstract}
We study the transport properties of a quantum dot-Majorana hybrid
system, in which each of paired Majorana bound states is connected
to one quantum dot. With the help of non-equilibrium Green's
function method, we obtain an exact solution of the Green's
functions and calculate the currents through the quantum dots and
nonlocal noise cross correlation between the currents. As a function
of dot energy levels $\epsilon_{1}$ and $\epsilon_{2}$, we find that
for the symmetric level configuration $\epsilon_{1}=\epsilon_{2}$,
the noise cross correlation is negative in the low lead voltage
regime, while it becomes positive with the increase of the lead
voltages. Due to the particle-hole symmetry, the cross correlation
is always positive in the anti-symmetric case
$\epsilon_{1}=-\epsilon_{2}$. In contrast, the cross correlation of
non-Majorana setups is always positive. For comparison, we also
perform the diagonalized master equation calculation to check its
applicability. It is found that the diagonalized master equations
work well in most regimes of system parameters. Nevertheless, it
shows an obvious deviation from the exact solution by the
non-equilibrium Green's function method when all eigenenergies of
the dot-Majorana hybrid system and simultaneously the energy
intervals are comparable to the dot-lead coupling strength.
\end{abstract}


\date{\today}

\maketitle

\section{Introduction}

Majorana fermions, defined as fermions equivalent to own
antiparticles, have been being hunted by high energy physicists for
a long time.\cite{Majorana1937,Nayak08RMP,Stern10Nat} Recent years,
the search of Majorana fermions has been shifted to solid-state
systems, such as in the fractional quantum Hall system and $p$-wave
superconductors.\cite{Nayak08RMP,Stern10Nat,Ivanov01PRL,Kitaev03AP,Moore91NPB,Flensberg11PRL,Fu10PRL}
In particular, the Majorana bound states (MBSs) are predicted to
appear at two ends of a semiconductor nanowire, in the proximity of
an $s$-wave superconductor and under a proper magnetic
field.\cite{Sau10PRL,Alicea11NP,Jiang11PRL,Bonderson11PRL,Oreg10PRL,Lutchyn10PRL,Alicea10PRB,Sau10PRB,Beenakker13ARCMP}
The signatures for possible formation of a spatially separated pair
of MBSs were reported in several
experiments.\cite{Mourik12Science,Deng12NL,Das12NP,Rokhinson12NP,Finck13PRL}
Two well-separated MBSs can define a nonlocal fermion level and its
occupation encodes a
qubit.\cite{Kitaev03AP,Moore91NPB,Flensberg11PRL} This nonlocal
topological qubit is immune to local perturbation and thus, has
potential application in quantum information. However, for the same
reason, it is of great challenge to coherently transfer and to read
out the quantum information of the topological
states.\cite{Flensberg11PRL,Beenakker13ARCMP,Leijnse11PRL} It has
been suggested that the MBS-quantum dot hybrid system might be one
of the solutions to the
problem.\cite{Flensberg11PRL,Leijnse11PRL,Leijnse12PRB} Up to now,
various MBS-dot hybrid
devices\cite{Leijnse11PRL,Leijnse12PRB,Tewari08PRL,Liu11PRB,Leijnse11PRB,Wang13PRB,Cao12PRB,Lu12PRB,Zocher13PRL,Liu13PRB,Cao12arXiv,Wang13EPL,Sau12NC}
have been proposed to detect existence of
MBS,\cite{Liu11PRB,Wang13PRB,Cao12PRB} to modulate nonlocal
correlation,\cite{Lu12PRB,Zocher13PRL,Liu13PRB,Cao12arXiv,Wang13EPL}
to estimate lifetime,\cite{Leijnse11PRB} and to remove the effect of
disorder.\cite{Sau12NC} Therefore, it is important to investigate
the transport properties of the MBS-dot hybrid systems.

One of the fascinating properties of MBSs is that MBSs could induce
nonlocal current cross correlation when they are coupled to
mesoscopic
circuits.\cite{Bolech07PRL,Law09PRL,Nilsson08PRL,Wu12PRB,Komnik09PRB,Joho12PRB}
It has been shown that a positive cross correlation could be induced
when MBSs couple to electron reservoirs
directly.\cite{Bolech07PRL,Law09PRL,Nilsson08PRL} In our previous
work, we proposed a device to modulate the Majorana-fermion-mediated
cross correlation of the currents flowing through two quantum dots
located between the MBSs and reservoirs.\cite{Lu12PRB} Subsequently,
the time-dependent evaluation of electron population and other
related transport properties were investigated by the scattering
matrix theory\cite{Zocher13PRL,Liu13PRB} and the master equation
approach,\cite{Cao12arXiv,Wang13EPL} respectively.

In this paper, we systematically investigate the nonlocal transport
properties of the quantum dot-Majorana hybrid systems (as shown in
Fig. \ref{fig:setup}) by means of both the non-equilibrium Green's
function (NEGF) method and diagonalized master equation (DME)
approach, respectively. The currents through the quantum dots and
nonlocal cross-correlation between the currents are investigated as
functions of lead voltage, dot energy levels, Majorana energy
splitting, dot-Majorana coupling, and temperature. In the weak
dot-lead coupling regime, we obtain the analytical expressions of
the currents and their noise correlation. The sign distribution of
the cross-correlation is analyzed (summarized in Table
\ref{tab:S12}), which shows a distinct 4-region feature from pure
quantum-dot or superconducting circuits (see Figs. \ref{fig:E0} and
\ref{fig:E0-noMBS} for comparison). Furthermore, in the
noninteracting case, the NEGF method gives the exact solution,
offering a benchmark to check the applicability of DME. The paper is
organized as follows. In Sec. \ref{sec:model}, the model Hamiltonian
of the dot-Majorana hybrid device, as well as the current and noise
cross correlation formulas in terms of NEGF and DME, are introduced.
The details of derivation are given in Appendices. In Sec.
\ref{sec:results}, the numerical results for the current and noise
cross correlation are presented. In Sec. \ref{sec:DME}, the
applicability of the DME approach is discussed by comparing with the
exact solution by the NEGF method. Finally, a summary is given in
Sec. \ref{sec:summary}.

\begin{table}[htbp]
\centering \caption{Sign distribution of noise correlation $S_{12}$
with the increase of lead voltage $V_{0}$ in the weak dot-lead
coupling regime. For simplicity, we consider the case of symmetric
Majorana-dot and dot-lead coupling strength, i.e.,
$\Gamma_{1}=\Gamma_{2}$ and $\lambda_{1}=\lambda_{2}$.}

\label{tab:S12} %
\begin{tabular}{c|c|c|c}
\hline
Cases  & Small $V_{0}$  & \ \ $\Rightarrow$ \ \  & Large $V_{0}$\tabularnewline
\hline
$\epsilon_{1}>0$, $\epsilon_{2}>0$  & $\mathrm{-}$  & $\mathrm{-}$  & + \tabularnewline
\hline
$\epsilon_{1}>0$, $\epsilon_{2}<0$  & +  & +  & + \tabularnewline
\hline
$\epsilon_{1}<0$, $\epsilon_{2}<0$  & $\mathrm{-}$  & +  & + \tabularnewline
\hline
$\epsilon_{1}<0$, $\epsilon_{2}>0$  & +  & +  & + \tabularnewline
\hline
\end{tabular}
\end{table}

\section{Model and formulism\label{sec:model}}

\subsection{Model}

\begin{figure}[htbp]
\centering \includegraphics[width=0.45\textwidth]{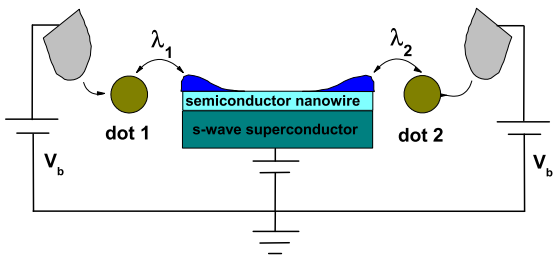} \caption{Schematic of a quantum dot-Majorana hybrid system. Two quantum dots
are coupled to two ends of a semiconductor nanowire with strong spin-orbit
interaction. The nanowire is in contact with an $s$-wave superconductor.
In a large enough Zeeman field, the nanowire is driven into the topological
superconducting phase and a pair of Majorana bound states appears
at its ends. Each dot is connected with a normal metal lead.}
\label{fig:setup}
\end{figure}

The schematic of the setup is illustrated in Fig. \ref{fig:setup}.
Two quantum dots are connected to a one-dimensional semiconductor
nanowire with strong Rashba spin-orbit
coupling.\cite{Tewari08PRL,Lu12PRB} Under a proper magnetic field,
the semiconductor nanowire resembles a topological superconductor
when it is adjacent to an $s$-wave superconductor due to the
proximity
effect.\cite{Sau10PRL,Alicea11NP,Jiang11PRL,Bonderson11PRL,Oreg10PRL,Lutchyn10PRL,Alicea10PRB,Sau10PRB,Beenakker13ARCMP}
At the ends of the wire, a pair of MBSs emerges. Each dot is
connected to one of the MBSs and one normal metal lead.
Experimentally, the magnetic field is up to several Tesla, which can
produce a Zeeman splitting in each dot as well such that the spin
degree of freedom of electrons can be suppressed. Consequently the
model Hamiltonian is expressed as
\begin{eqnarray}
H & = & H_{0}+H_{L}+H_{T}.
\end{eqnarray}
The quantum dot-Majorana hybrid part is described by
\begin{eqnarray}
H_{0} & = & \sum_{j=1,2}\epsilon_{j}d_{j}^{\dagger}d_{j}+\frac{i}{2}\epsilon_{M}\eta_{a}\eta_{b}\notag\\
 &  & +\lambda_{1}(d_{1}^{\dagger}-d_{1})\eta_{a}+i\lambda_{2}(d_{2}^{\dagger}+d_{2})\eta_{b},
\end{eqnarray}
where $\epsilon_{j}$ is the energy level in quantum dot $j$
($j=1,2$) and $d_{j}(d_{j}^{\dag})$ is the annihilation (creation)
operator of electron. The quantum dots $j$ are coupled to the MBSs
with the strength $\lambda_{j}$. $\epsilon_{Mj}$ is the coupling
strength between two MBSs $\eta_{ja}(=\eta_{ja}^{\dag})$ and
$\eta_{jb}(=\eta_{jb}^{\dag})$. The electron reservoirs and their
coupling to the dots are described by the Hamiltonian,
\begin{eqnarray}
H_{L} & = & \sum\limits _{jk}\epsilon_{j,k}c_{jk}^{\dagger}c_{jk},\nonumber \\
H_{T} & = & \sum\limits _{j,k}(t_{j}c_{jk}^{\dagger}d_{j}+h.c.),
\end{eqnarray}
where $\epsilon_{i,k}$ is the electron energy in the lead $i$ and
$t_{i}$ is the tunneling amplitude.

For convenience, we can switch from the Majorana fermion
representation to the regular fermion representation by defining
$\eta_{a}=f+f^{\dagger}$, $\eta_{b}=i(f^{\dagger}-f)$ with
$f^{\dagger}f-ff^{\dagger}=1$.\cite{Tewari08PRL,Liu11PRB,Lu12PRB} In
the new representation, the Hamiltonian in the central region
becomes\cite{Lu12PRB}
\begin{eqnarray}
H_{0} & = & \sum_{j=1,2}\epsilon_{j}d_{j}^{\dagger}d_{j}+\epsilon_{M}(f^{\dagger}f-\frac{1}{2})\notag\\
 &  & +[\lambda_{1}(f^{\dagger}d_{1}+fd_{1})+\lambda_{2}(f^{\dagger}d_{2}-fd_{2})+h.c.].
\end{eqnarray}

\subsection{Current and noise cross correlation}

The operator of tunneling current from the lead $\alpha$ to the
central region can be given by
\begin{eqnarray}
\mathbf{I}_{\alpha}(t) & \equiv & -e\frac{d\hat{N}_{\alpha}}{dt}=\frac{ie}{\hbar}[\hat{N}_{\alpha},\hat{H}]\notag\\
 & = & \frac{ie}{\hbar}t_{\alpha}\sum_{k}(c_{\alpha k}^{\dagger}d_{\alpha}-d_{\alpha}^{\dagger}c_{\alpha k}).
\end{eqnarray}
The current noise correlations are defined as
\begin{eqnarray}
S_{12}(t-t')=\langle I_{1}(t)I_{2}(t')+I_{2}(t')I_{1}(t)\rangle-2\langle I_{1}\rangle\langle I_{2}\rangle.
\end{eqnarray}
$S_{12}$ is referred to as the noise cross correlation between the
currents flowing through dot 1 and dot 2.

The quantum transport of many-body systems can be obtained by
several techniques. In the following subsections, we will calculate
the currents and their noise cross correlations using both the
standard Keldysh NEGF method\cite{Meir92PRL,Jauho94PRB,Sun05PRB} and
the DME
approach.\cite{Lambert07PRB,Kashcheyevs06PRB,Mitra07PRB,Timm08PRB,Poltl09PRB,Lu09PRB}

\subsection{Non-equilibrium Green's function method}

Following Ref. \onlinecite{Sun05PRB}, we define the matrix of lesser
Green's function $\mathbf{G}^{<}$ in the Nambu space spanned by the
spinor
$\psi=(d_{1},d_{2},f,c_{1},c_{2},d_{1}^{\dagger},d_{2}^{\dagger},f^{\dagger},c_{1}^{\dagger},c_{2}^{\dagger})^{T}$
, where $c_{i}$ is the electron annihilation operator in lead $i$
and
\begin{equation}
\mathbf{G}^{<}(t-t')=\langle\langle\psi(t)|\psi^{\dagger}(t')\rangle\rangle^{<}=i\langle\psi^{\dagger}(t')\psi(t)\rangle.
\end{equation}
In the frequency space,
\begin{equation}
\mathbf{G}_{\omega}^{<}=\int dt\mathbf{G}^{<}(t)e^{i\omega t}
\end{equation}
In this representation, the currents are given by
\begin{eqnarray}
I_{\alpha}=\mathbf{Tr}\left[\mathbf{\hat{I}}_{\alpha}\mathbf{G}^{<}\right],
\end{eqnarray}
and the noise spectrum $S_{12}(\omega)$ is given by
\begin{eqnarray}
S_{12}(\omega) & = & \int d\omega'\mathbf{Tr}\left[\mathbf{\hat{I}}_{1}
\mathbf{G}_{\omega'}^{<}\mathbf{\hat{I}}_{2}\mathbf{G}_{\omega'+\omega}^{>}
+\mathbf{\hat{I}}_{2}\mathbf{G}_{\omega'}^{>}\mathbf{\hat{I}}_{1}
\mathbf{G}_{\omega'+\omega}^{<}\right].\nonumber \\
\end{eqnarray}
The matrices of the current operators are given by
\begin{eqnarray}
\mathbf{I}_{1}=\frac{e}{2\hbar}t_{1}\left(\begin{array}{llllllll}
1 & 0\\
0 & 1
\end{array}\right)\otimes\left(\begin{array}{cccccc}
0 & 0 & 0 & -1 & 0\\
0 & 0 & 0 & 0 & 0\\
0 & 0 & 0 & 0 & 0\\
1 & 0 & 0 & 0 & 0\\
0 & 0 & 0 & 0 & 0
\end{array}\right),\nonumber \\
\mathbf{I}_{2}=\frac{e}{2\hbar}t_{2}\left(\begin{array}{llllllll}
1 & 0\\
0 & 1
\end{array}\right)\otimes\left(\begin{array}{cccccc}
0 & 0 & 0 & 0 & 0\\
0 & 0 & 0 & 0 & -1\\
0 & 0 & 0 & 0 & 0\\
0 & 0 & 0 & 0 & 0\\
0 & 1 & 0 & 0 & 0
\end{array}\right).
\end{eqnarray}
To find the lesser Green's function $\mathbf{G}^{<}$, we need to
calculate the retarded Green's function $\mathbf{G}^{r}$. As there
is no four-operator interaction term in the Hamiltonian, the Green's
function can be solved analytically. Performing the standard
equation of motion procedure for the central region, the retarded
Green's function ${\bf G}^{r}$ can be found and written in terms of
the Dyson equation ${\bf G}^{r}={\bf g}^{r}+{\bf g}^{r}{\bf
\Sigma}^{r}{\bf G}^{r}$,
\begin{equation}
\mathbf{G^{r}}=\left(1-\mathbf{g}^{r}\Sigma^{r}\right)^{-1}\mathbf{g}^{r}.
\end{equation}
Here ${\bf g}^{r}$ is the bare Green's function of the central
region when isolated from the leads (\textit{i.e.}, when
$t_{1}=t_{2}=0$),
\begin{eqnarray}
 &  & {\bf g}^{r}(\omega)^{-1}\nonumber \\
 & = & \left(\begin{array}{llllllll}
1 & 0\\
0 & 1
\end{array}\right)\otimes\left(\begin{array}{llllllll}
\omega & 0 & 0 & 0 & 0\\
0 & \omega & 0 & 0 & 0\\
0 & 0 & \omega & 0 & 0\\
0 & 0 & 0 & i/\pi\rho & 0\\
0 & 0 & 0 & 0 & i/\pi\rho
\end{array}\right)\nonumber \\
 &  & -\left(\begin{array}{llllllll}
1 & 0\\
0 & -1
\end{array}\right)\otimes\left(\begin{array}{llllllll}
\epsilon_{1} & 0 & \lambda_{1} & 0 & 0\\
0 & \epsilon_{2} & \lambda_{2} & 0 & 0\\
\lambda_{1} & \lambda_{2} & \epsilon_{M} & 0 & 0\\
0 & 0 & 0 & 0 & 0\\
0 & 0 & 0 & 0 & 0
\end{array}\right)\nonumber \\
 &  & +\left(\begin{array}{llllllll}
0 & 1\\
-1 & 0
\end{array}\right)\otimes\left(\begin{array}{llllllll}
0 & 0 & -\lambda_{1} & 0 & 0\\
0 & 0 & \lambda_{2} & 0 & 0\\
\lambda_{1} & -\lambda_{2} & 0 & 0 & 0\\
0 & 0 & 0 & 0 & 0\\
0 & 0 & 0 & 0 & 0
\end{array}\right),
\end{eqnarray}
and the self-energy ${\bf \Sigma}^{r}$ has the form
\begin{eqnarray}
{\bf \Sigma}^{r}=\left(\begin{array}{llllllll}
1 & 0\\
0 & -1
\end{array}\right)\otimes\left(\begin{array}{llllllll}
0 & 0 & 0 & t_{1} & 0\\
0 & 0 & 0 & 0 & t_{2}\\
0 & 0 & 0 & 0 & 0\\
t_{1} & 0 & 0 & 0 & 0\\
0 & t_{2} & 0 & 0 & 0
\end{array}\right).
\end{eqnarray}
Using the relationship $\mathbf{G}^{r}=(\mathbf{G}^{a})^{\dagger}$,
the advanced Green's function can be found. It is straightforward to
have the lesser Green's function from the standard Keldysh equation,
\begin{eqnarray}
\mathbf{G}^{<} & = & (1+\mathbf{G}^{r}\mathbf{\Sigma}^{r})\mathbf{g}^{<}
(1+\mathbf{\Sigma}^{a}\mathbf{G}^{a})+\mathbf{G}^{r}\mathbf{\Sigma}^{<}
\mathbf{G}^{a}\nonumber \\
 & = & \mathbf{G}^{r}\mathbf{g}^{r-1}\mathbf{g}^{<}\mathbf{g}^{a-1}
 \mathbf{G}^{a}+\mathbf{G}^{r}\mathbf{\Sigma}^{<}\mathbf{G}^{a}.
\end{eqnarray}
In the present case, $\mathbf{\Sigma}^{<}=0$ and
\begin{eqnarray}
\mathbf{g}^{r-1}\mathbf{g}^{<}\mathbf{g}^{a-1}=\left(\begin{array}{cccccc}
\mathbf{\Lambda}_{e} & 0\\
0 & \mathbf{\Lambda}_{h}
\end{array}\right),
\end{eqnarray}
with
\begin{eqnarray}
\mathbf{\Lambda}_{e(h)}=\left(\begin{array}{llllllll}
0 & 0 & 0 & 0 & 0\\
0 & 0 & 0 & 0 & 0\\
0 & 0 & 0 & 0 & 0\\
0 & 0 & 0 & \frac{2i}{\pi\rho}f(\omega\mp\mu_{1}) & 0\\
0 & 0 & 0 & 0 & \frac{2i}{\pi\rho}f(\omega\mp\mu_{2})
\end{array}\right),
\end{eqnarray}
where $f(\omega)=[1+e^{\omega/k_{B}T}]^{-1}$ is the Fermi-Dirac
distribution function and $k_{B}T$ is the temperature.

\subsection{Diagonalized master equations}

The electronic transport through this system in a sequential
tunneling regime can also be described by the quantum rate equations
for dynamical evolution of the density matrix
elements.\cite{Kashcheyevs06PRB,Mitra07PRB,Timm08PRB,Poltl09PRB,Lu09PRB}
In this subsection we present the formula of the currents and their
cross correlation in the diagonalized
representation.\cite{Poltl09PRB} In comparison with the NEGF method
and the regular master equation approach, the DME approach is
convenient in the calculation of many interacting and strongly
coherent energy levels.

In the dot-Majorana hybrid part, there are eight states in the
particle-number representation $|n_{1}n_{2}n_{M}\rangle$, where
$n_{i}=0,1$ is the occupation numbers in dot $i$ and the MBS. By
choosing the basis $\{|000\rangle$, $|101\rangle$, $|011\rangle$,
$|110\rangle$, $|001\rangle$, $|100\rangle$, $|010\rangle$,
$|111\rangle\}$, the eigen equation for the dot-Majorana Hamiltonian
is given by
\begin{eqnarray}
\left(\begin{array}{cccccc}
\mathbf{M}^{e} & 0\\
0 & \mathbf{M}^{o}
\end{array}\right)\Psi_{i}=E_{i}\Psi_{i},
\end{eqnarray}
where
\begin{eqnarray}
\mathbf{M}^{e}=\left(\begin{array}{cccccc}
-\frac{\epsilon_{M}}{2} & \lambda_{1} & -\lambda_{2} & 0\\
\lambda_{1} & \epsilon_{1}+\frac{\epsilon_{M}}{2} & 0 & \lambda_{2}\\
-\lambda_{2} & 0 & \epsilon_{2}+\frac{\epsilon_{M}}{2} & -\lambda_{1}\\
0 & \lambda_{2} & -\lambda_{1} & \epsilon_{1}+\epsilon_{2}-\frac{\epsilon_{M}}{2}
\end{array}\right),
\end{eqnarray}
\begin{eqnarray}
\mathbf{M}^{o}=\left(\begin{array}{cccccc}
\frac{\epsilon_{M}}{2} & \lambda_{1} & \lambda_{2} & 0\\
\lambda_{1} & \epsilon_{1}-\frac{\epsilon_{M}}{2} & 0 & -\lambda_{2}\\
\lambda_{2} & 0 & \epsilon_{2}-\frac{\epsilon_{M}}{2} & -\lambda_{1}\\
0 & -\lambda_{2} & -\lambda_{1} & \epsilon_{1}+\epsilon_{2}+\frac{\epsilon_{M}}{2}
\end{array}\right).
\end{eqnarray}
As the matrix is block diagonalized, we denote the eigenstates in
terms of
$\Psi_{l}^{o(e)}=(a_{l}^{o(e)},b_{l}^{o(e)},c_{l}^{o(e)},d_{l}^{o(e)})^{\mathbf{T}}$
for
\begin{equation}
\mathbf{M}^{o(e)}\Psi_{l}^{o(e)}=E_{l}^{o(e)}\Psi_{l}^{o(e)}.
\end{equation}

In the DME approach the density matrix
$\rho_{D}(t)=|\Psi_{i}\rangle\langle\Psi_{i'}|$ has only diagonal
terms that represent the populations in the states $\Psi_{i}^{o(e)}$
and its time evaluation is governed by the rate equation
\begin{eqnarray}
\frac{d}{dt}\rho_{D}(t)=\mathbf{W}\rho_{D}(t),
\end{eqnarray}
where the elements of the rate matrix are given
by\cite{Lu12PRB,Poltl09PRB}
\begin{eqnarray}
W_{k'k} & = & \sum_{i}\Gamma_{i}\left[f(\Delta_{k'k}+\mu_{i})|\langle\beta_{k'}|d_{i}|\beta_{k}\rangle|^{2}\right.\nonumber \\
 &  & \left.+f(\Delta_{k'k}-\mu_{i})|\langle\beta_{k'}|d_{i}^{\dagger}|\beta_{k}\rangle|^{2}\right]
\end{eqnarray}
for $k\neq k'$ and
\begin{eqnarray}
W_{kk} & = & -\sum_{k'\neq k}^{N}W_{k'k}.
\end{eqnarray}
Here$\mu_{i}$ is the chemical potential in lead $i$, and
$\Delta_{k'k}$ is the Bohr frequency of the transition from
$|\beta_{k}\rangle$ to $|\beta_{k'}\rangle$. In the wide-band limit
approximation, the coupling between dot $i$ and its lead is denoted
by $\Gamma_{i}=2\pi|t_{i}|^{2}\rho_{i}$ with $\rho_{i}$ the spinless
density of states near the Fermi surface of lead $i$.

The steady-state current $I_{i}$ is given by
\begin{eqnarray}
I_{i} & = & e\sum_{k}[\hat{\mathbf{\Gamma}}^{i}\rho_{D}^{(0)}]_{k},
\end{eqnarray}
where $\hat{\mathbf{\Gamma}}^{i}$ is the matrix of the current
operator and its element is given by
\begin{eqnarray}
\hat{\Gamma}_{k'k}^{i} & = & \Gamma_{i}\left[f(\Delta_{k'k}+\mu_{i})|\langle\beta_{k'}|d_{i}|\beta_{k}\rangle|^{2}\right.\nonumber \\
 &  & \left.-f(\Delta_{k'k}-\mu_{i})|\langle\beta_{k'}|d_{i}^{\dagger}|\beta_{k}\rangle|^{2}\right].
\end{eqnarray}
The first term of $\hat{\Gamma}_{k'k}^{i}$ is the tunneling current
flowing into the lead and the second term is the tunneling current
flowing from the lead to dot.

We are interested in the cross correlation induced by the MBS
between the currents through the quantum dots. The noise power
spectra can be expressed as the Fourier transform of the
current-current correlation function
\begin{eqnarray}
S_{I_{i}I_{j}}(\omega) & = & 2\langle I_{i}(t)I_{j}(0)\rangle_{\omega}-2\langle I_{i}\rangle_{\omega}\langle I_{j}\rangle_{\omega},
\end{eqnarray}
where $I_{i}$ is the current in dot $i$ and $t$ is the time.
Furthermore, the current-current correlation function can be
expressed in the ${\bm{\rho}}$ representation as follows
\begin{eqnarray}
\langle{I_{i}(t)I_{j}(0)}\rangle & = & \theta(t)\sum_{k}[\hat{\Gamma}^{i}\hat{T}(t)\hat{\Gamma}^{j}{\bm{\rho}}^{(0)}]_{k}\nonumber \\
 &  & +\theta(-t)\sum_{k}[\hat{\Gamma}^{j}\hat{T}(-t)\hat{\Gamma}^{j}{\bm{\rho}}^{(0)}]_{k}
\end{eqnarray}
with $\hat{T}(t)=\exp[\mathbf{W}t]$ the propagator governing the
time evaluation of the density matrix element $\rho_{k}(t)$.
Finally, the current-current correlation in the $\omega$-space
becomes
\begin{eqnarray}
\langle I_{i}(t)I_{j}(0)\rangle_{\omega} & = &
\sum_{k}\left[\hat{\Gamma}^{i}\hat{T}(\omega)\hat{\Gamma}^{j}
{\bm{\rho}}^{(0)}+\hat{\Gamma}^{j}\hat{T}(-\omega)\hat{\Gamma}^{i}
{\bm{\rho}}^{(0)}\right]_{k},\nonumber \\
\end{eqnarray}
where $\hat{T}(\pm\omega)=\left(\mp
i\omega\hat{I}-\mathbf{W}\right)^{-1}$.

\section{Currents and noise cross correlation\label{sec:results}}

In the section, we will calculate the current and zero-frequency
cross correlation $S_{12}$ as functions of the lead voltage,
Majorana energy splitting, dot energy levels, and temperature. The
results by both the NEGF method and the DME approach will be
presented simultaneously for comparison. In most calculations, we
use the same dot-lead tunneling rate and dot-Majorana coupling
strength for both dots, \emph{i.e.}, $\Gamma_{i}=\Gamma_{0}$ and
$\lambda_{i}=\lambda_{0}$, while the effects of asymmetric dot-lead
and dot-Majorana coupling are discussed in Sec.
\ref{sec:asymmetric}. All energies and frequencies are measured in
units of $\Gamma_{0}$. The voltages $V_{0}=\mu_{1}=\mu_{2}$ are
symmetrically applied to both leads.

\begin{figure}[htbp]
\centering \includegraphics[width=0.45\textwidth]{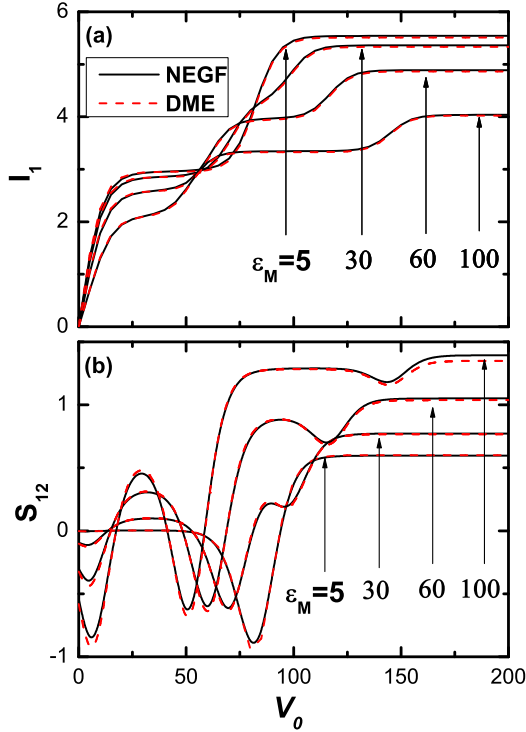} \caption{The current $I_{1}$ through dot 1 (top) and zero-frequency cross
correlation $S_{12}$ (bottom) as functions of the voltage $V_{0}$
in the leads, for different Majorana energy splitting $\epsilon_{M}=5$,
$30$, $60$, and $100$, respectively. The results obtained from
the non-equilibrium Green's function (NEGF) method and the diagonalized
master equation (DME) approach are compared. Other parameters: $k_{B}T=5$,
$\lambda_{0}=40$, $\epsilon_{1}=20$, and $\epsilon_{2}=25$, where
the dot-lead coupling strength $\Gamma_{0}$ is taken as the energy
unit.}
\label{fig:V0}
\end{figure}

\begin{figure}[htbp]
\centering \includegraphics[width=0.45\textwidth]{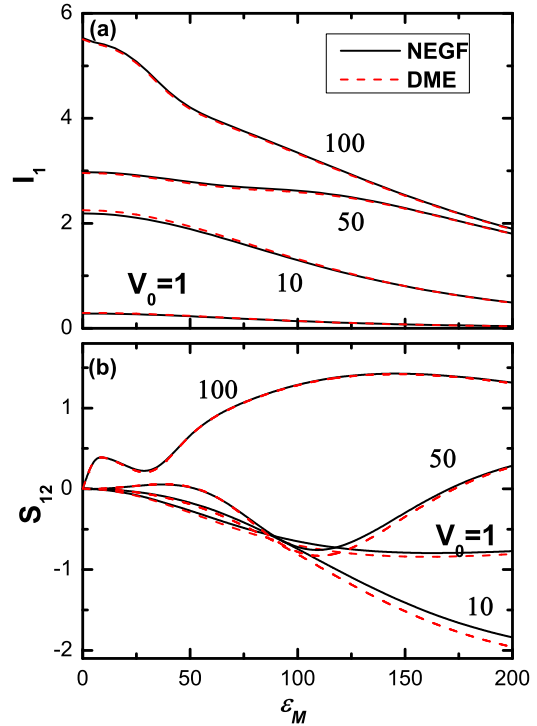} \caption{The current $I_{1}$ (top) and zero-frequency cross correlation $S_{12}$
(bottom) as functions of $\epsilon_{M}$ for different lead voltages
$V_{0}=1$, $10$, $50$, and $100$, respectively. Other parameters
are $k_{B}T=5$, $\lambda_{0}=40$, $\epsilon_{1}=20$, and $\epsilon_{2}=25$.}
\label{fig:EM}
\end{figure}

\subsection{Tuning lead voltage }

In Fig. \ref{fig:V0} we show the charge current $I_{1}$ and cross
correlation $S_{12}$ as functions of the voltage $V_{0}$ for
different values of the Majorana energy splitting $\epsilon_{M}$. As
shown in Fig. \ref{fig:V0}, the currents and cross correlations
obtained from the two methods agree very well with each other. At
low temperatures, the I-V curve exhibits the standard staircase
behavior, ascending in steps and rising to a higher plateau every
time the chemical potential of the leads $\mu_{i}$ crosses a higher
energy level of the central region. For low voltage $\mu_{i}$,
electrons in the leads do not have enough energy to tunnel into the
dots, and the currents are contributed by thermal excitations. As
the chemical potential of the lead exceeds the eigenenergies one by
one, the transport channels are opened correspondingly. Fig.
\ref{fig:V0} considers the case that both dot energy levels are
higher than the middle position of the superconducting gap. At a
relatively low voltage, the cross correlation favors a negative
correlation, corresponding to the competitive relationship between
two tunneling paths. With the increase of voltage, the cross
correlation shows a sign reversal and becomes positive, indicating
that the correlation between two channels becomes cooperative.

\subsection{Effect of Majorana energy splitting}

The Majorana energy splitting $\epsilon_{M}$ between two MBSs is a
key parameter to generate nonlocal cross correlation and the
dependence of the cross correlation on $\epsilon_{M}$ is shown in
Fig. \ref{fig:EM}. It has been demonstrated
theoretically\cite{Sarma12PRB} and experimentally\cite{Finck13PRL}
that the splitting $\epsilon_{M}$ has an oscillatory dependence on
the Zeeman field and the chemical potential in the nanowire, and
exponentially decays as a function of the wire length. For a
nanowire with its length $\sim1\mu$m, the typical energy range of
$\epsilon_{M}$ is about $0-50\mu$eV.\cite{Sarma12PRB} In the limit
of $\epsilon_{M}=0$, there is no overlap between the wave functions
of two Majorana's bound states. The interaction between dot $i$ and
MBS on one side changes the parity of the MBS, but does not affect
the dot-Majorana interaction on the other side. In this case, there
is no communication between two MBSs, corresponding to a zero
$S_{12}$. It has been pointed out that when $\epsilon_{M}$
approaches to zero, the cross correlation $S_{12}$ is proportional
to $\epsilon_{M}^{2}$ and thus vanishes at the point
$\epsilon_{M}=0$.\cite{Zocher13PRL} We show that the results are
confirmed by both NEGF and DME calculations. With the increase of
$\epsilon_{M}$, the current $I_{1}$ is suppressed to zero gradually
due to the detuning of the dot-Majorana interaction. At relatively
low voltage $V_{0}$, the lowest eigenenergy level will be shifted
out of the transport window if $\epsilon_{M}$ is large enough.
Correspondingly, the cross correlation $S_{12}$ is governed by the
thermal noise for $\epsilon_{M}\gg V_{0}$.

\subsection{Tuning dot energy levels }

One of the roles of the quantum dots is to control the nonlocal
cross correlation by tuning the dot energy levels, for potential
applications in the Majorana-based topological quantum
information.\cite{Lu12PRB} Experimentally, typical values are a
dot-lead coupling $\Gamma_i$ in the order of 1$\mu$eV, which is much
weaker than other system parameters. In the weak dot-lead coupling
limit, we find the explicit expressions of the current and noise
cross correlation (the details can be found in Appendix
\ref{sec:I-S}). For the symmetric level configuration
($\epsilon_0=\epsilon_{1}=\epsilon_{2}$), and in the limit of large
lead voltage $V_{0}$, we have
\begin{eqnarray}
S_{12}\approx\int d\omega\frac{16\epsilon_{M}^{2}\lambda_{0}^{4}
\Gamma_{0}^{2}(\epsilon_{0}^{2}-\omega^{2})^{2}}{\left[A_\omega+B_\omega^{2}/A_\omega\right]^{2}},
\label{S12-sym-largeV}
\end{eqnarray}
where $A_\omega$ and $B_\omega$ are defined as
\begin{eqnarray}
A_\omega & = & (\omega^{2}-\epsilon_{M}^{2})(\epsilon_{0}^{2}-\omega^{2})^{2}
+8\lambda_{0}^{2}\omega^{2}(2\lambda_{0}^{2}+\epsilon_{0}^{2}-\omega^{2}),
\nonumber \\
B_\omega & = & 2\Gamma_{0}\omega\left[(\omega^{2}-4\lambda_{0}^{2}
-\epsilon_{0}^{2})(\omega^{2}-2\lambda_{0}^{2}-\epsilon_{M}^{2})
-4\lambda_{0}^{2}\epsilon_{M}^{2}\right].
\nonumber\\
\end{eqnarray}
It can be seen that $S_{12}$ is always positive in this case. In the
small $V_{0}$ limit,
\begin{eqnarray}
S_{12}\approx\int d\omega\frac{16\epsilon_{M}^{2}\lambda_{0}^{4}\Gamma_{0}^{2}
\left[(4\epsilon_{0}\omega)^{2}[(f_{\omega}-1)f_{\omega}]\right]}
{\left[A_\omega+B_\omega^{2}/A_\omega\right]^{2}},\label{S12-sym-smallV}
\end{eqnarray}
where the Fermi function $f_{\omega}\in[0,1]$, then $S_{12}$ is
always negative. Moreover, one can check in Eq.
(\ref{S12-symmetric}) that
$S_{12}(\epsilon_{0})<S_{12}(-\epsilon_{0})$ for $\epsilon_{0}>0$
and any $V_{0}$. Therefore, with the increase of the lead voltage,
$S_{12}$ in the regime $\epsilon_{0}<0$ experiences a sign reversal
from negative to positive. As $V_{0}$ increases further, the sign
reversal of $S_{12}$ then occurs in the regime $\epsilon_{0}>0$.

For the anti-symmetric level configuration
($\epsilon_0=\epsilon_1=-\epsilon_2$), in the large lead voltage
limit,
\begin{eqnarray}
S_{12}\approx\int d\omega\frac{16\epsilon_{M}^{2}\lambda_{0}^{4}
\Gamma_{0}^{2}\left[(\epsilon_{0}^{2}-\omega^{2})^{2}+8\epsilon_{0}^{2}
\omega^{2}\right]}{\left[A_\omega+B_\omega^{2}/A_\omega\right]^{2}},\label{S12-asym-largeV}
\end{eqnarray}
and in the small voltage limit
\begin{eqnarray}
S_{12}\approx\int d\omega\frac{16\epsilon_{M}^{2}\lambda_{0}^{4}
\Gamma_{0}^{2}\left[8\epsilon_{0}^{2}\omega^{2}(f_{\omega}^{+}
-f_{\omega}^{-})^{2}\right]}{\left[A_\omega+B_\omega^{2}/A_\omega\right]^{2}}.\label{S12-asym-smallV}
\end{eqnarray}
Therefore, $S_{12}$ is always positive in the anti-symmetric level
configuration. Different from the symmetric case, $S_{12}$ is an
even function of $\epsilon_{0}$ for the anti-symmetric level
configuration, due to the particle-hole symmetry.

\begin{figure}[htbp]
\centering
\includegraphics[width=0.48\textwidth]{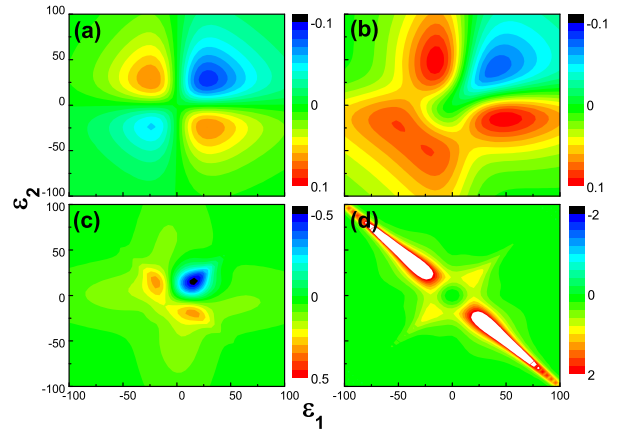}
\caption{The zero-frequency cross correlation $S_{12}$
obtained from the NEGF method as functions of dot energy levels $\epsilon_{1}$ and $\epsilon_{2}$
for different voltages (a) $V_{0}=1$, (b) $10$, (c) $30$, and (d) $200$,
respectively. Other parameters are $k_{B}T=5$, $\lambda_{0}=20$,
and $\epsilon_{M}=8$.}
\label{fig:E0}
\end{figure}

\begin{figure}[htbp]
\centering
\includegraphics[width=0.48\textwidth]{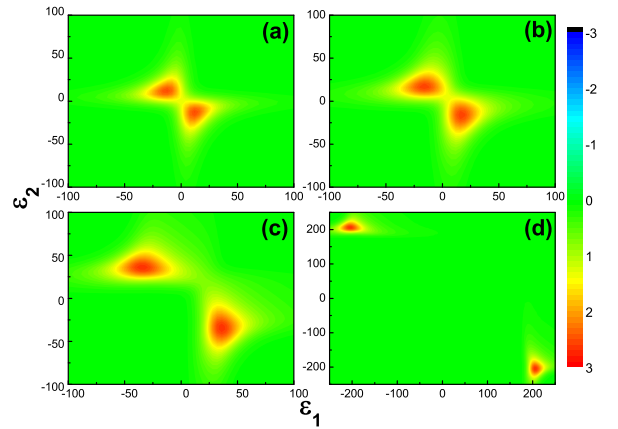}
\caption{The same as Fig. \ref{fig:E0}, except that the MBSs are replaced by a normal $s$-wave superconductor.}
\label{fig:E0-noMBS}
\end{figure}

From the above discussion, we could conclude the sign of $S_{12}$ as
a function of dot energy levels in the limit of weak dot-lead
coupling, as presented in Table \ref{tab:S12}. In Fig. \ref{fig:E0}
we further demonstrate the cross correlation $S_{12}$ obtained from
the NEGF method as functions of the dot energy levels $\epsilon_{i}$
for different voltages $V_{0}$, which is consistent with the
analytical results. For the small lead voltage comparable to the
dot-lead coupling strength, a four peaked clover-like pattern of
noise cross correlation is illustrated when modulating the dot
energy levels. In the previous experiments, such a four peaked
clover-like pattern is also demonstrated in a capacitively-coupled
double dot device, where the sign of the cross correlation is
independent on the lead voltage.\cite{McClure07PRL,Zhang07PRL} For
the present Majorana-dot hybrid device, the lead voltage determines
which eigenlevel participates in the transport. With the increase of
$V_0$, $S_{12}$ in the region of $\epsilon_{1,2}<0$ becomes positive
firstly and is much weaker than other regions. For large enough
$V_0$, the transport channels through all eigenlevels are open and
their interaction leads the positive $S_{12}$ in all regions. In
this case, the cross correlation $S_{12}$ in the region of
antisymmetric level configuration $\epsilon_1=-\epsilon_2$, is much
stronger than that for the symmetric case.

The signature of Majorana fermions in the cross correlation is
different from those of non-Majorana setups. In our setup, the key
feature of the cross correlation is the four peaked clover-like
pattern as a function of the dot levels (as illustrated in Fig.
\ref{fig:E0}), due to crossed Andreev reflections mediated by the
MBSs. In contrast, the sign distribution is absent in the
experiments of other normal superconductor-quantum dot hybrid
systems.\cite{Hofstetter09Nature,Hofstetter11PRL,Wei12NatPhys,Das12NatCommun}
Also, a setup with a normal s-wave superconductor was theoretically
studied, but there the cross correlation was not
addressed.\cite{Eldridge10PRB} We calculate the cross correlation of
this non-Majorana setup for comparison (the details can be found in
Appendix \ref{sec:non-Majorana}). As shown in Fig.
\ref{fig:E0-noMBS}, the clover-like cross correlation distribution
in Fig. \ref{fig:E0} is absent in the non-Majorana setup, where the
cross correlation is always positive and the peaks appear at
$\epsilon_1=-\epsilon_2=\pm V_0$. The comparison of Figs.
\ref{fig:E0} and \ref{fig:E0-noMBS} gives a distinguishable
signature of the MBSs in the cross correlation spectrum.

\begin{figure}[htbp]
\centering \includegraphics[width=0.48\textwidth]{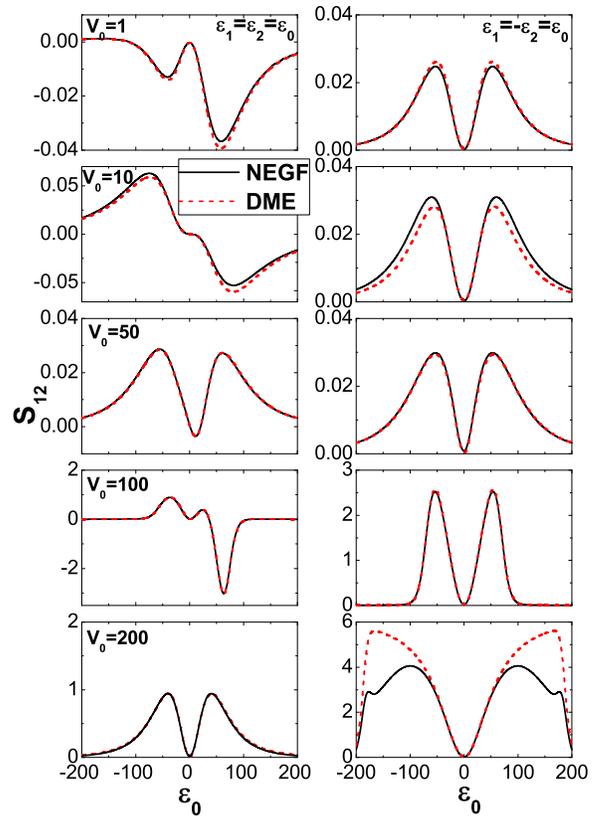} \caption{The
zero-frequency cross correlation $S_{12}$ as functions of (left)
symmetric dot energy levels $\epsilon_{0}(=\epsilon_{1}=\epsilon_{2})$
and (right) anti-symmetric dot energy levels $\epsilon_{0}(=\epsilon_{1}=-\epsilon_{2})$
for different voltages $V_{0}=1$, $10$, $50$, $100$, and $200$,
respectively. Other parameters are $k_{B}T=5$, $\lambda_{0}=40$,
and $\epsilon_{M}=10$.}
\label{fig:E0-2}
\end{figure}

In previous studies, the applicability of the DME approach in
Majorana-dot hybrid systems is questioned, especially as a function
of dot energy level.\cite{Zocher13PRL} Above we have demonstrated in
a wide range of system parameters that the DME approach could give
the numerical results with high accuracy. In principle, there is no
extra limitation on the configuration of the dot energy levels when
the DME approach is applied. To confirm this point, we present the
cross correlation $S_{12}$ calculated by the NEGF method and the DME
approach in Fig.  \ref{fig:E0-2}. We separately consider the
symmetric ($\epsilon_{1}=\epsilon_{2}$) and anti-symmetric
($\epsilon_{1}=-\epsilon_{2}$) cases of dot level configurations. In
the symmetric configuration ($\epsilon_{1}=\epsilon_{2}$), $S_{12}$
shows a strong dependence on the voltage in the leads. For
relatively small voltage $V_{0}$, $S_{12}$ exhibits a pattern of
negative values. With the increase of $V_{0}$, the sign reversal of
$S_{12}$ appears as a function of $\epsilon_{0}$. It should be noted
$S_{12}$ experiences twice sign reversals in the region
$\epsilon_0>0$, due to the competition between tunneling pathes
flowing through different eigenlevels. For large $V_{0}$, $S_{12}$
becomes positive eventually. In the anti-symmetric level
configuration ($\epsilon_{1}=-\epsilon_{2}$), $S_{12}$ vanishes at
$\epsilon_{i}=0$ and shows two peaks lying on both sides of
$\epsilon_{i}=0$. In this case, $S_{12}$ is always positive for
different configurations of dot energy levels. Due to the
particle-hole symmetry, $S_{12}$ should be symmetric as a function
of $\epsilon_{0}$ when changing the sign of $\epsilon_{0}$. This
symmetry can be captured by both the NEGF method and the DME
approach, as shown in Fig. \ref{fig:E0-2}.

\subsection{Effect of temperature}

\begin{figure}[htbp]
\centering \includegraphics[width=0.48\textwidth]{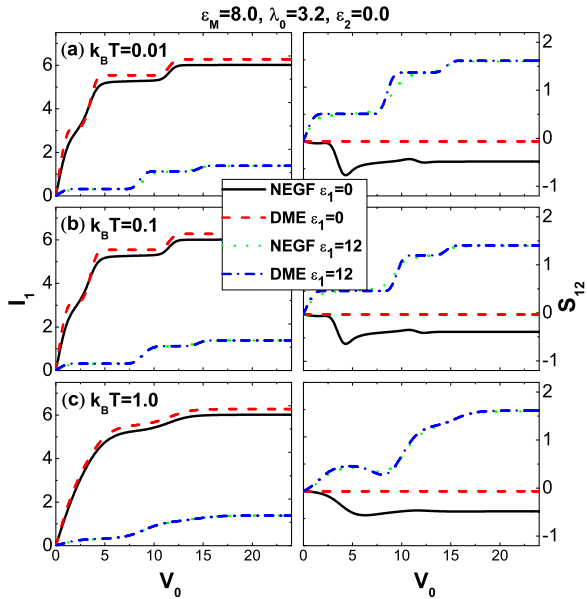} \caption{The
current $I_{1}$ (left) and zero-frequency cross correlation $S_{12}$
(right) as functions of lead voltage $V_{0}$ for different values
of temperature $k_{B}T=0.01$, $0.1$, and $1$. Two level configurations
are examined. One is $(\epsilon_{1},\epsilon_{2})=(0,0)$ and the
other is $(\epsilon_{1},\epsilon_{2})=(12,0)$. Other parameters:
$\lambda_{0}=3.2$ and $\epsilon_{M}=8$, which is same to the parameters
used in Fig. 2 (b) of Ref. \onlinecite{Zocher13PRL}.}

\label{fig:temperature}
\end{figure}

\begin{figure}[htbp]
\centering \includegraphics[width=0.48\textwidth]{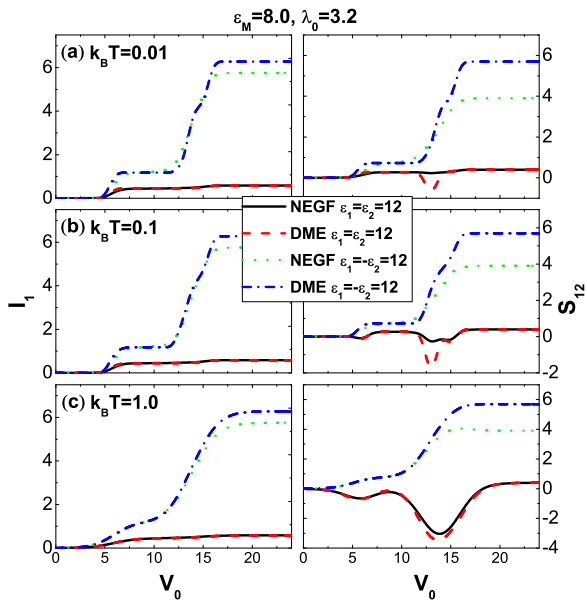} \caption{The
current $I_{1}$ (left) and zero-frequency cross correlation $S_{12}$
(right) as functions of lead voltage $V_{0}$ for different values
of temperature $k_{B}T=0.01$, $0.1$, and $1$. The symmetric level
configuration $\epsilon_{1}=\epsilon_{2}$ and anti-symmetric case
$\epsilon_{1}=-\epsilon_{2}$ are considered. Other parameters: $\lambda_{0}=3.2$
and $\epsilon_{M}=8$, which is same to the parameters used in Fig.
2 (d) of Ref. \onlinecite{Zocher13PRL}.}

\label{fig:temperature-2}
\end{figure}

Figures \ref{fig:temperature} and \ref{fig:temperature-2} show the
current $I_{1}$ and cross correlation $S_{12}$ as functions of the
voltage $V_{0}$ for different temperatures $k_{B}T$. For comparison,
we use the parameters from the previous work (Fig. 2 of Ref.
\onlinecite{Zocher13PRL}). The effect of thermal fluctuation is
reflected in the Fermi distributions in both leads. At a low
temperature, e.g., $k_{B}T=0.01$, the I-V curves exhibit the
staircases and the current jumps to another step when the voltage
$V_{0}$ crosses one of the eigenenergies of the dot-MBS Hamiltonian.
The DME approach assumes the dot-lead coupling as perturbation. In
the sequential tunneling regime, the broadening effect in the
conductance by the dot-lead coupling is not included in the DME
approach. Compared to the results by the NEGF method, the I-V curves
by the DME approach display steeper steps at low temperatures. With
the increase of temperature, more electrons away from the Fermi
energy contribute to the current and thermal fluctuation-induced
tunneling is enhanced. For $k_{B}T\gg\Gamma_{0}$, the temperature
effect becomes dominant. Therefore, one of the applicable
requirements of the DME approach is that the temperature is higher
than the dot-lead coupling strength.

At the low temperature $k_{B}T=0.01$, the results by the NEGF method
agree well with those by the scattering matrix
theory.\cite{Zocher13PRL} It is noted that for the case
$\epsilon_{1}=\epsilon_{2}=0$, although the currents calculated by
the NEGF method and the DME approach agree well with each other,
$S_{12}$ shows a large deviation. For different $k_{B}T$, $S_{12}$
by the DME approach always approaches zero, while the NEGF method
gives the negative noise cross correlation. However, for other three
dot level configurations, $S_{12}$ by the NEGF method and the DME
approach are in good agreement, even at low temperatures. This
implies that the DME approach fails in the calculation of
current-current correlation when all the eigenenergies and their
interval are comparable to the dot-lead coupling strength.

\begin{figure}[htbp]
\centering \includegraphics[width=0.5\textwidth]{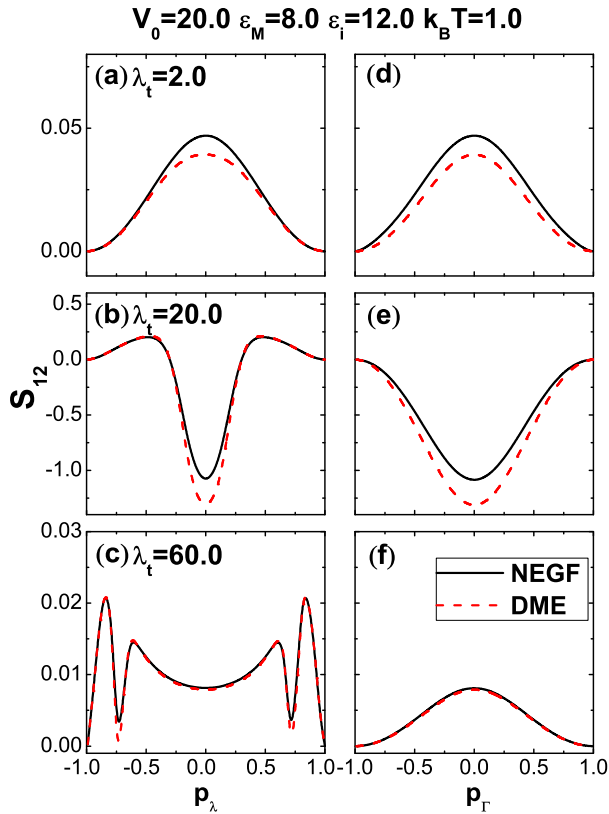} \caption{The cross correlation $S_{12}$ as functions of {[}(a)-(c){]} dot-Majorana
coupling asymmetry $p_{\lambda}$ and {[}(d)-(f){]} dot-lead coupling
asymmetry $p_{\Gamma}$ for different $\lambda_{t}=\lambda_{1}+\lambda_{2}=2.0$,
$20.0$, and $60.0$, respectively. Other parameters are $k_{B}T=1.0$,
$\epsilon_{M}=8.0$, $\epsilon_{i}=12.0$, $V_{0}=20.0$, and $\Gamma_{1}+\Gamma_{2}=2.0$.}

\label{fig:asymmetry}
\end{figure}

\subsection{Asymmetric dot-Majorana and dot-lead coupling\label{sec:asymmetric}}

In the above discussion we assume a symmetric coupling scenario,
\emph{i.e.}, $\Gamma_{1}=\Gamma_{2}$ and $\lambda_{1}=\lambda_{2}$.
Actually, the dot-Majorana and dot-lead couplings determine the
dwell time of an electron through the dots, so asymmetric coupling
could also modulate the currents and their correlations. To
characterize the asymmetry, we define the parameters
\begin{eqnarray}
p_{\lambda}=\frac{\lambda_{1}-\lambda_{2}}{\lambda_{1}+\lambda_{2}}
\end{eqnarray}
for the dot-Majorana coupling, and
\begin{eqnarray}
p_{\Gamma}=\frac{\Gamma_{1}-\Gamma_{2}}{\Gamma_{1}+\Gamma_{2}}
\end{eqnarray}
for the dot-lead coupling. Figure 4 presents the cross correlation
$S_{12}$ as functions of $p_{\lambda}$ and $p_{\Gamma}$. As shown in
Fig. \ref{fig:asymmetry}, the results by two methods show slight
discrepancy for weak dot-Majorana coupling strength
$\lambda_{t}=\lambda_{1}+\lambda_{2}$. The dot-Majorana interaction
$\lambda_{i}$ is the energy scale that measures the relaxation rate
of the central region. In the presence of strong asymmetry
($p_{\lambda}\rightarrow\pm1$ and $p_{\Gamma}\rightarrow\pm1$), the
cross correlation $S_{12}$ is suppressed considerably. For strong
dot-Majorana or lead-dot coupling asymmetry, an electron needs more
time to tunnel into or out of one of the double dots, resulting in
the suppression of the current flowing through the dot and cross
correlation. Figure \ref{fig:asymmetry} (b) shows that for strong
dot-Majorana interaction, the coupling asymmetry $p_{\lambda}$ not
only modulates the strength of $S_{12}$, but also could reverse the
sign of $S_{12}$.

\begin{figure}[htbp]
\centering \includegraphics[width=0.48\textwidth]{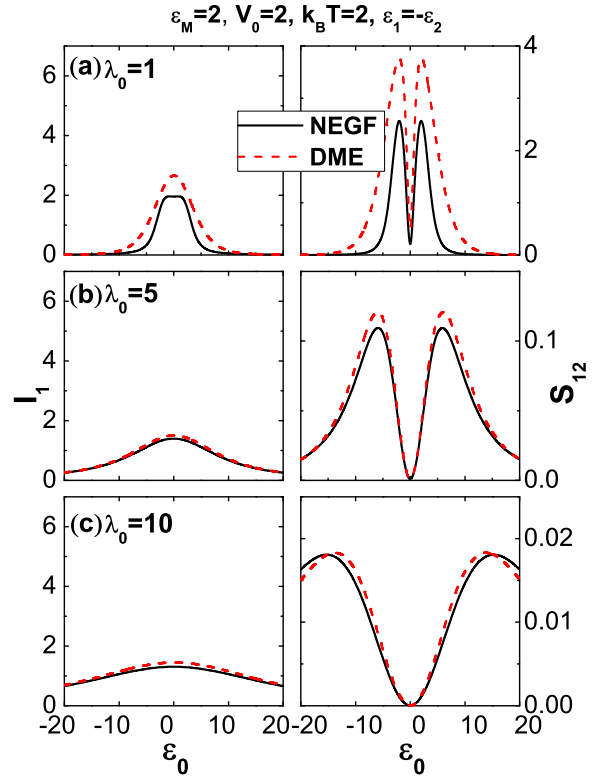} \caption{
The current $I_{1}$ (left) and zero-frequency cross correlation $S_{12}$
(right) as functions of anti-symmetric dot energy level
$\epsilon_{0}(=\epsilon_{1}=-\epsilon_{2})$
for different voltages $\lambda_{0}=1$, $5$, and $10$, respectively.
Other parameters: $k_{B}T=2$, $\epsilon_{M}=2$, and $V_{0}=2$.}

\label{fig:DME}
\end{figure}

\section{Applicability of diagonalized master equations\label{sec:DME}}

The NEGF method and DME approach are two common tools widely used in
mesoscopic tunneling
systems.\cite{Bruus04Book,Zimbovskayaa11PR,Levitov03Book,Blanter00PR,Kashcheyevs06PRB,Mitra07PRB,Timm08PRB,Poltl09PRB,Lu09PRB,Lambert07PRB,Pulido08PRB}
Although the applicability of the NEGF method and the DME approach
have been addressed
elsewhere,\cite{Kashcheyevs06PRB,Poltl09PRB,White13JCP,Volkovich11PCCP}
it is still unclear for unconventional fermionic systems, such as
Majorana fermion system. For this MBS-dot hybrid device, it is of
great interest whether the DME method could correctly reflect the
information of current and noise, as questioned in a previous
work.\cite{Zocher13PRL}

In the previous sections we demonstrated in a wide range of system
parameters that the DME approach could give the numerical results
with high accuracy. However, one still needs to be careful in two
cases. The first is that the DME approach assumes that the dot-lead
coupling much weaker than other system energy
scales.\cite{Poltl09PRB} We demonstrate the importance of this
restriction in Fig. \ref{fig:DME}, in which the coherence strength
$\lambda_{0}$ takes values of $\Gamma_{0}$, $5\Gamma_{0}$, and
$10\Gamma_{0}$ and $\epsilon_{M}(=2\Gamma_{0})$ is comparable to
$\Gamma_{0}$. It is shown that when all the eigenenergies of $H_{0}$
and their intervals are comparable to the dot-lead coupling
strength, the results of the DME approach shows a clear deviation
from those of the NEGF method. The other case is when degenerate
eigenstates appear, the matrix for the stationary state population
is linearly relevant and DME is not appropriate. This leads to the
obvious deviation between the results of the NEGF method and the DME
approach which appears at relatively large voltages for the
anti-symmetric level configuration $\epsilon_{1}=-\epsilon_{2}$ in
Figs. \ref{fig:E0-2} and \ref{fig:temperature}. For
$\epsilon_{1}=-\epsilon_{2}$, there exists energy degeneracies in
the eigenstates, as a result of particle-hole symmetry.

\section{Summary\label{sec:summary}}

In summary, we investigated the currents and their cross correlation
in a double dot-Majorana hybrid system by using the NEFG method and
the DME approach. We systematically studies the effects of the dot
energy levels, Majorana energy splitting, Majorana-dot interaction,
and the chemical potential in the leads on the transport properties.
In the weak dot-lead coupling regime, the analytical expressions of
currents and noise cross correlation are presented. It is found that
for the symmetric dot level configuration
$\epsilon_{1}=\epsilon_{2}$, the noise cross correlation is negative
in the small lead voltage case. With the increase of lead voltage,
the cross correlation in the cases of $\epsilon_{i}<0$ and
$\epsilon_{i}>0$ turns to positive successively. In contrast, the
cross correlation is always positive and symmetric about zero dot
level in the anti-symmetric case $\epsilon_{1}=-\epsilon_{2}$, as a
result of particle-hole symmetry. In addition, the sign diagram of
the noise cross correlation is presented as a function of dot energy
levels. In the present system, the NEGF method can give the exact
solution of the transport properties, which provides a benchmark to
check the applicability of the DME approach. The results obtained by
both methods agree reasonably well in most cases. However, the
results of the DME approach show a clear deviation in the cross
correlation for strong the dot-lead coupling strength or when energy
degeneracy appears in the central region. The comparison will be a
reference when generalizing DME to larger systems with many more
dot-Majorana units.

\begin{acknowledgments}
This work was supported by the Research Grants Council, University
Grants Committee, Hong Kong under Grant No. HKU 703713P and National
Natural Science Foundation of China under Grants No. 61474018.
\end{acknowledgments}

\begin{widetext}

\appendix

\section{For non-equilibrium Green's function method\label{sec:NEGF}}

In the main context, we present a compact form of NEGF to deduce the
formula of tunneling currents and their correlations, which is
equivalent to the result of the scattering matrix approach in the
noninteracting case. In this appendix, we will give another set of
standard formula of NEGF that has been widely used in previous
studies.\cite{Bruus04Book,Meir92PRL,Jauho94PRB,Fang07JPCM} This
version of NEGF could take the many-body interaction into account in
the proceeding of the equation of motion truncation. In the
noninteracting case, it is equivalent to NEGF in the main context.
First we introduce the lesser Green's functions, which are defined
as
\begin{eqnarray}
\langle\langle c_{\alpha k}(t)|d_{\beta}^{\dagger}(t')\rangle\rangle^{<} & = & i\langle d_{\beta}^{\dagger}(t')c_{\alpha k}(t)\rangle,\nonumber \\
\langle\langle d_{\alpha}(t)|c_{\beta k}^{\dagger}(t')\rangle\rangle^{<} & = & i\langle c_{\beta k}^{\dagger}(t')d_{\alpha}(t)\rangle.
\end{eqnarray}
In terms of the lesser Green's functions, the current is written
into
\begin{eqnarray}
I_{\alpha}(t)=\frac{e}{\hbar}t_{\alpha}\sum_{k}[\langle\langle
d_{\alpha}(t)|c_{\alpha k}^{\dagger}(t)\rangle\rangle^{<}
-\langle\langle c_{\alpha k}(t)|d_{\alpha}^{\dagger}(t)\rangle\rangle^{<}],
\end{eqnarray}
and the cross correlation $S_{12}$ could be expressed as
\begin{eqnarray}
S_{12}(t-t') & = & \frac{2e^{2}}{\hbar^{2}}t_{1}t_{2}\sum_{k,k'}
\left[\langle\langle d_{2}(t')|c_{1k}^{\dagger}(t)\rangle\rangle^{<}
\langle\langle c_{2k'}^{\dagger}(t')|d_{1}(t)\rangle\rangle^{<}
+\langle\langle c_{2k'}(t')|d_{1}^{\dagger}(t)\rangle\rangle^{<}
\langle\langle d_{2}^{\dagger}(t')|c_{1k}(t)\rangle\rangle^{<}\right.
\nonumber \\
&  & \left.+\langle\langle c_{2k'}^{\dagger}(t')|d_{1}^{\dagger}(t)
\rangle\rangle^{<}\langle\langle d_{2}(t')|c_{1k}(t)\rangle\rangle^{<}
+\langle\langle d_{2}^{\dagger}(t')|c_{1k}^{\dagger}(t)\rangle\rangle^{<}
\langle\langle c_{2k'}(t')|d_{1}(t)\rangle\rangle^{<}\right.
\nonumber \\
&  & \left.-\langle\langle c_{2k'}^{\dagger}(t')|c_{1k}^{\dagger}(t)
\rangle\rangle^{<}\langle\langle d_{2}(t')|d_{1}(t)\rangle\rangle^{<}
-\langle\langle c_{2k'}(t')|c_{1k}(t)\rangle\rangle^{<}\langle\langle
d_{2}^{\dagger}(t')|d_{1}^{\dagger}(t)\rangle\rangle^{<}\right.
\nonumber \\
&  & \left.-\langle\langle c_{2k'}^{\dagger}(t')|c_{1k}(t)\rangle
\rangle^{<}\langle\langle d_{2}(t')|d_{1}^{\dagger}(t)\rangle\rangle^{<}
-\langle\langle c_{2k'}(t')|c_{1k}^{\dagger}(t)\rangle\rangle^{<}
\langle\langle d_{2}^{\dagger}(t')|d_{1}(t)\rangle\rangle^{<}\right].
\end{eqnarray}
In the above equations, we have approximately truncate the
four-operator correlation functions,\cite{Fang07JPCM} \emph{e.g}.,
\begin{eqnarray}
\langle c_{\alpha k}^{\dagger}d_{\alpha}c_{\beta k'}^{\dagger}d_{\beta}
\rangle & \approx & \langle c_{\alpha k}^{\dagger}d_{\beta}\rangle\langle
d_{\alpha}c_{\beta k'}^{\dagger}\rangle-\langle c_{\alpha k}^{\dagger}
c_{\beta k'}^{\dagger}\rangle\langle d_{\alpha}d_{\beta}\rangle+\langle
c_{\alpha k}^{\dagger}d_{\alpha}\rangle\langle c_{\beta k'}^{\dagger}d_{\beta}\rangle,
\nonumber \\
\langle c_{\alpha k}^{\dagger}d_{\alpha}d_{\beta}^{\dagger}c_{\beta k'}
\rangle & \approx & \langle c_{\alpha k}^{\dagger}c_{\beta k'}\rangle\langle
d_{\alpha}d_{\beta}^{\dagger}\rangle-\langle c_{\alpha k}^{\dagger}
d_{\beta}^{\dagger}\rangle\langle d_{\alpha}c_{\beta k'}\rangle
+\langle c_{\alpha k}^{\dagger}d_{\alpha}\rangle\langle d_{\beta}^{\dagger}c_{\beta k'}\rangle.
\end{eqnarray}
After applying a Fourier transform, the stationary current can be
expressed as
\begin{eqnarray}
I_{\alpha} & = & \frac{e}{\hbar}\int d\omega\left[t_{\alpha}
\langle\langle d_{\alpha}|c_{\alpha k}^{\dagger}\rangle\rangle^{<}
-t_{\alpha}\langle\langle c_{\alpha k}|d_{\alpha}^{\dagger}\rangle\rangle^{<}\right],
\end{eqnarray}
and the cross correlation spectrum is given by
\begin{eqnarray}
S_{12}(\omega) & = & \int dtS_{12}(t)e^{i\omega t}=\frac{4\pi e^{2}}
{\hbar^{2}}t_{1}t_{2}\int d\omega'\sum_{k,k'}\left[\langle\langle
d_{2}|c_{1k}^{\dagger}\rangle\rangle_{\omega'+\omega}^{<}\langle\langle
c_{2k'}^{\dagger}|d_{1}\rangle\rangle_{\omega'}^{<}+\langle\langle
c_{2k'}|d_{1}^{\dagger}\rangle\rangle_{\omega'+\omega}^{<}\langle\langle
d_{2}^{\dagger}|c_{1k}\rangle\rangle_{\omega'}^{<}\right.
\nonumber \\
&  & \left.+\langle\langle c_{2k'}^{\dagger}|d_{1}^{\dagger}\rangle
\rangle_{\omega'+\omega}^{<}\langle\langle d_{2}|c_{1k}\rangle
\rangle_{\omega'}^{<}+\langle\langle d_{2}^{\dagger}|c_{1k}^{\dagger}
\rangle\rangle_{\omega'+\omega}^{<}\langle\langle c_{2k'}|d_{1}\rangle
\rangle_{\omega'}^{<}-\langle\langle c_{2k'}^{\dagger}|c_{1k}^{\dagger}
\rangle\rangle_{\omega'+\omega}^{<}\langle\langle d_{2}|d_{1}\rangle
\rangle_{\omega'}^{<}\right.
\nonumber \\
&  & \left.-\langle\langle c_{2k'}|c_{1k}\rangle\rangle_{\omega'}^{<}
 \langle\langle d_{2}^{\dagger}|d_{1}^{\dagger}\rangle\rangle_{\omega'
 +\omega}^{<}-\langle\langle c_{2k'}^{\dagger}|c_{1k}\rangle
 \rangle_{\omega'}^{<}\langle\langle d_{2}|d_{1}^{\dagger}\rangle
 \rangle_{\omega'+\omega}^{<}-\langle\langle c_{2k'}|c_{1k}^{\dagger}
 \rangle\rangle_{\omega'+\omega}^{<}\langle\langle d_{2}^{\dagger}
 |d_{1}\rangle\rangle_{\omega'}^{<}\right],
\end{eqnarray}
where the Green's functions involving both dot and lead operators
can be readily related to the Green's functions of only dot
operators and only lead operators. By applying the Langreth analytic
continuation rules\cite{Meir92PRL,Jauho94PRB}
\begin{eqnarray}
\langle\langle d_{\alpha}^{(\dagger)}|c_{\beta k}^{\dagger}\rangle\rangle^{<}
& = & t_{\beta}\left[\langle\langle d_{\alpha}^{(\dagger)}|d_{\beta}^{\dagger}
\rangle\rangle^{r}\langle\langle c_{\beta k}|c_{\beta k}^{\dagger}\rangle
\rangle^{0<}\right.\left.+\langle\langle d_{\alpha}^{(\dagger)}|d_{\beta}
^{\dagger}\rangle\rangle^{<}\langle\langle c_{\beta k}|c_{\beta k}^{\dagger}
\rangle\rangle^{0a}\right],
\nonumber \\
\langle\langle d_{\alpha}^{(\dagger)}|c_{\beta k}\rangle\rangle^{<} & = &
-t_{\beta}\left[\langle\langle d_{\alpha}^{(\dagger)}|d_{\beta}\rangle\rangle^{r}
\langle\langle c_{\beta k}^{\dagger}|c_{\beta k}\rangle\rangle^{0<}\right.
\left.+\langle\langle d_{\alpha}^{(\dagger)}|d_{\beta}\rangle\rangle^{<}\langle
\langle c_{\beta k}^{\dagger}|c_{\beta k}\rangle\rangle^{0a}\right],
\nonumber \\
\langle\langle c_{\beta k}^{\dagger}|d_{\alpha}^{(\dagger)}\rangle\rangle^{<}
& = & t_{\beta}\left[\langle\langle d_{\beta}^{\dagger}|d_{\alpha}^{(\dagger)}
\rangle\rangle^{r}\langle\langle c_{\beta k}^{\dagger}|c_{\beta k}\rangle
\rangle^{0<}\right.\left.+\langle\langle d_{\beta}^{\dagger}|d_{\alpha}^{(\dagger)}
\rangle\rangle^{<}\langle\langle c_{\beta k}^{\dagger}|c_{\beta k}
\rangle\rangle^{0a}\right],\nonumber \\
\langle\langle c_{\beta k}|d_{\alpha}^{(\dagger)}\rangle\rangle^{<} & = &
-t_{\beta}\left[\langle\langle d_{\beta}|d_{\alpha}^{(\dagger)}\rangle\rangle^{r}
\langle\langle c_{\beta k}|c_{\beta k}^{\dagger}\rangle\rangle^{0<}\right.
\left.+\langle\langle d_{\beta}|d_{\alpha}^{(\dagger)}\rangle\rangle^{<}
\langle\langle c_{\beta k}|c_{\beta k}^{\dagger}\rangle\rangle^{0a}\right],
\end{eqnarray}
where $\langle\langle c_{\alpha k}|c_{\beta
k}^{\dagger}\rangle\rangle^{0}$ is the bare Green's functions for
the leads. For the present model, the exotic contact terms
$\langle\langle c_{\alpha k}|c_{\beta k'}\rangle\rangle^{<}$ and
$\langle\langle c_{\alpha k}^{\dagger}|c_{\beta
k'}^{\dagger}\rangle\rangle^{<}$ are also kept due to the pairing
effect induced by MBSs and they could be obtained from dot-lead
Green's functions
\begin{eqnarray}
\langle\langle c_{\alpha k}^{(\dagger)}|c_{\beta k'}^{\dagger}\rangle\rangle^{<}
& = & t_{\beta}\left[\langle\langle c_{\alpha k}^{(\dagger)}|d_{\beta}^{\dagger}
\rangle\rangle^{r}\langle\langle c_{\beta k'}|c_{\beta k'}^{\dagger}\rangle
\rangle^{0<}\right.\left.+\langle\langle c_{\alpha k}^{(\dagger)}|d_{\beta}^{\dagger}
\rangle\rangle^{<}\langle\langle c_{\beta k'}|c_{\beta k'}^{\dagger}\rangle
\rangle^{0a}\right],\nonumber \\
\langle\langle c_{\alpha k}^{(\dagger)}|c_{\beta k'}\rangle\rangle^{<} & = &
-t_{\beta}\left[\langle\langle c_{\alpha k}^{(\dagger)}|d_{\beta}\rangle
\rangle^{r}\langle\langle c_{\beta k'}^{\dagger}|c_{\beta k'}\rangle\rangle^{0<}
\right.\left.+\langle\langle c_{\alpha k}^{(\dagger)}|d_{\beta}\rangle
\rangle^{<}\langle\langle c_{\beta k'}^{\dagger}|c_{\beta k'}\rangle
\rangle^{0a}\right].
\end{eqnarray}
The bare Green's functions for the leads are given by
\begin{eqnarray}
\langle\langle c_{\alpha k}|c_{\alpha k}^{\dagger}\rangle\rangle_{\omega}^{0<} & = & i\pi f(\omega)\delta(\omega-\epsilon_{\alpha k}),\nonumber \\
\langle\langle c_{\alpha k}^{\dagger}|c_{\alpha k}\rangle\rangle_{\omega}^{0<} & = & i\pi[1-f(-\omega)]\delta(\omega+\epsilon_{\alpha k}),\nonumber \\
\langle\langle c_{\alpha k}|c_{\alpha k}^{\dagger}\rangle\rangle_{\omega}^{0r(a)} & = & \mp i\pi\delta(\omega-\epsilon_{\alpha k}),\nonumber \\
\langle\langle c_{\alpha k}^{\dagger}|c_{\alpha k}\rangle\rangle_{\omega}^{0r(a)} & = & \mp i\pi\delta(\omega+\epsilon_{\alpha k}).
\end{eqnarray}
Performing the standard equation of motion procedure for the central
region, we could obtain the retarded Green's functions in the Nambu
space spanned by
$\psi=(d_{1},d_{2},f,d_{1}^{\dagger},d_{2}^{\dagger},f^{\dagger})^{T}$,
\begin{eqnarray}
\mathbf{M}\mathbf{G}^{r}=1
\end{eqnarray}
where
\begin{eqnarray}
\mathbf{M}=\left(\begin{array}{cccccccc}
\omega-\epsilon_{1}+i\Gamma_{1} & 0 & -\lambda_{1} & 0 & 0 & -\lambda_{1}\\
0 & \omega-\epsilon_{2}+i\Gamma_{2} & -\lambda_{2} & 0 & 0 & \lambda_{2}\\
-\lambda_{1} & -\lambda_{2} & \omega-\epsilon_{M} & \lambda_{1} & -\lambda_{2} & 0\\
0 & 0 & \lambda_{1} & \omega+\epsilon_{1}+i\Gamma_{1} & 0 & \lambda_{1}\\
0 & 0 & -\lambda_{2} & 0 & \omega+\epsilon_{2}+i\Gamma_{2} & \lambda_{2}\\
-\lambda_{1} & \lambda_{2} & 0 & \lambda_{1} & \lambda_{2} & \omega+\epsilon_{M}
\end{array}\right)
\end{eqnarray}
and $G^{r}=\langle\langle\psi|\psi^{\dagger}\rangle\rangle^{r}$.

Using the relationship $\mathbf{G}^{r}=(\mathbf{G}^{a})^{\dagger}$,
we can obtain the advanced Green's function. The related lesser
Green's functions can be calculated by the Keldysh equation
$G^{<}=G^{r}\Sigma^{<}G^{a}$. In the present representation, the
self energy $\Sigma^{<}$ is
\begin{eqnarray}
\mathbf{\Sigma}^{<}(\omega)=\left(\begin{array}{cccccc}
\mathbf{\Sigma}_{e}^{<} & 0\\
0 & \mathbf{\Sigma}_{h}^{<}
\end{array}\right),
\end{eqnarray}
with
\begin{eqnarray}
\mathbf{\Sigma}_{e(h)}^{<}=\left(\begin{array}{cccccc}
i\Gamma_{1}f(\omega\mp\mu_{1}) & 0 & 0\\
0 & i\Gamma_{2}f(\omega\mp\mu_{2}) & 0\\
0 & 0 & 0
\end{array}\right),
\end{eqnarray}
where $f(\omega)=[1+e^{\omega/k_{B}T}]^{-1}$ is the Fermi
distribution function and $k_{B}T$ is the system temperature.

\section{For Diagonalized master equation approach\label{sec:DME-derive}}

For the central region, we can diagonalize the Hamiltonian
$H_{0}=H_{D}+H_{M}+H_{DM}$ by solving their eigenenergy and the
corresponding eigenstates. For every state, we have
\begin{eqnarray}
H_{0}|000\rangle & = & -\frac{\epsilon_{M}}{2}|000\rangle+\lambda_{1}|101\rangle-\lambda_{2}|011\rangle,\nonumber \\
H_{0}|101\rangle & = & (\epsilon_{1}+\frac{\epsilon_{M}}{2})|101\rangle+\lambda_{1}|000\rangle+\lambda_{2}|110\rangle,\nonumber \\
H_{0}|011\rangle & = & (\epsilon_{2}+\frac{\epsilon_{M}}{2})|011\rangle-\lambda_{1}|110\rangle-\lambda_{2}|000\rangle,\nonumber \\
H_{0}|110\rangle & = & (\epsilon_{1}+\epsilon_{2}-\frac{\epsilon_{M}}{2})|110\rangle-\lambda_{1}|011\rangle+\lambda_{2}|101\rangle.
\end{eqnarray}
It can be seen that the above four states $|000\rangle$,
$|101\rangle$, $|011\rangle$, and $|110\rangle$ forms a closed block
of even parity. Similarly, another four states $|001\rangle$,
$|100\rangle$, $|010\rangle$, and $|111\rangle$ forms a closed block
of odd parity and they meets
\begin{eqnarray}
H_{0}|001\rangle & = & \frac{\epsilon_{M}}{2}|001\rangle+\lambda_{1}|100\rangle+\lambda_{2}|010\rangle,\nonumber \\
H_{0}|100\rangle & = & (\epsilon_{1}-\frac{\epsilon_{M}}{2})|101\rangle+\lambda_{1}|001\rangle-\lambda_{2}|111\rangle,\nonumber \\
H_{0}|010\rangle & = & (\epsilon_{2}-\frac{\epsilon_{M}}{2})|011\rangle-\lambda_{1}|111\rangle+\lambda_{2}|001\rangle,\nonumber \\
H_{0}|111\rangle & = & (\epsilon_{1}+\epsilon_{2}+\frac{\epsilon_{M}}{2})|111\rangle-\lambda_{1}|010\rangle-\lambda_{2}|100\rangle.
\end{eqnarray}
It should also be noted that for the tunnel operators, the
second-quantization operators in Eq. (23) can be written in the
relevant many-body basis as
\begin{eqnarray}
d_{1} & = & |001\rangle\langle101|+|010\rangle\langle110|+|000\rangle\langle100|+|011\rangle\langle111|,\nonumber \\
d_{1}^{\dagger} & = & |101\rangle\langle001|+|110\rangle\langle010|+|100\rangle\langle000|+|111\rangle\langle011|,\nonumber \\
d_{2} & = & |001\rangle\langle011|-|100\rangle\langle110|+|000\rangle\langle010|-|101\rangle\langle111|,\nonumber \\
d_{2}^{\dagger} & = & |011\rangle\langle001|-|110\rangle\langle100|+|010\rangle\langle000|-|111\rangle\langle101|.
\end{eqnarray}
Based on these relationships, we could obtain the matrix form of the
current operators in Eq. (23) in the diagonalized representation.
Correspondingly, the statistical averaging of  an any time-dependent
operator $\hat{A}(t)$ becomes
\begin{eqnarray}
\langle\hat{A}(t)\rangle={\rm Tr}\{\hat{A}\rho\}=\sum_{k}[{\bm{A}}{\bm{\rho}}(t)]_{k}=\sum_{k}[{\bm{A}}{\bm{\rho}}^{(0)}]_{k},
\end{eqnarray}
where ${\bm{A}}$ is the matrix expression of the operator $\hat{A}$,
${\bm{\rho}}^{(0)}$ is the steady state solution of the rate
equations, and $[\rho]_{k}$ is the k-th element of the vector
$\rho$.

\section{Currents and cross correlation in weak dot-lead coupling limit\label{sec:I-S}}

For weak dot-lead coupling strength, the expressions of the current
and noise cross correlation can be found with the help of NEGF. For
simplicity, we consider the symmetric dot-leading and dot-Majorana
coupling strength, i.e., $\Gamma_i=\Gamma_0$ and
$\lambda_i=\lambda_0$. A symmetric lead voltages $V_i=V_0$ is
applied in both leads.

In the symmetric level configuration $\epsilon_{1}=\epsilon_{2}$,
\begin{eqnarray}
I_{i}\approx\int d\omega\frac{8\lambda_{0}^{4}\Gamma_{0}^{2}
\left[f_{\omega}^{-}-f_{\omega}^{+}\right]\left[(\epsilon_{M}^{2}
+\omega^{2})(\epsilon_{0}^{2}-\omega^{2})^{2}+8\lambda_{0}^{2}\omega^{2}
(2\lambda_{0}^{2}+\epsilon_{0}^{2}-\omega^{2})\right]}{A_\omega^{2}+B_\omega^{2}},
\end{eqnarray}
\begin{eqnarray}
S_{12}\approx\int d\omega\frac{16\epsilon_{M}^{2}\lambda_{0}^{4}\Gamma_{0}^{2}
\left[(\epsilon_{0}^{2}-\omega^{2})^{2}(f_{\omega}^{-}-f_{\omega}^{+})^{2}
+4\epsilon_{0}\omega[(\epsilon_{0}+\omega)^{2}(f_{\omega}^{-}-1)f_{\omega}^{-}
-(\epsilon_{0}-\omega)^{2}(f_{\omega}^{+}-1)f_{\omega}^{+}]\right]}
{\left[A_\omega+B_\omega^{2}/A_\omega\right]^{2}},\label{S12-symmetric}
\end{eqnarray}
where $f_{\omega}^{\pm}=\left[1+e^{(\omega\pm
V_{0})/k_{B}T}\right]^{-1}$,
\begin{eqnarray}
A_\omega & = & (\omega^{2}-\epsilon_{M}^{2})(\epsilon_{0}^{2}-\omega^{2})^{2}
+8\lambda_{0}^{2}\omega^{2}(2\lambda_{0}^{2}+\epsilon_{0}^{2}-\omega^{2}),
\nonumber \\
B_\omega & = & 2\Gamma_{0}\omega\left[(\omega^{2}-4\lambda_{0}^{2}
-\epsilon_{0}^{2})(\omega^{2}-2\lambda_{0}^{2}-\epsilon_{M}^{2})
-4\lambda_{0}^{2}\epsilon_{M}^{2}\right].
\end{eqnarray}
In the limit of large lead voltage, $f_{\omega}^{-}=1$,
$f_{\omega}^{+}=0$, then one obtains Eq. (\ref{S12-sym-largeV}). In
the limit of small lead voltage, $f_{\omega}^{-}\approx
f_{\omega}^{+}=0\equiv f_{\omega}$, then one obtains Eq.
(\ref{S12-sym-smallV}). Moreover, by using the relationship
$f_{\omega}^{-}=1-f_{-\omega}^{+}$, one can check that
$S_{12}(\epsilon_{0})<S_{12}(-\epsilon_{0})$ for $\epsilon_{0}>0$
for any $V_{0}$.

In the anti-symmetric level configuration
$\epsilon_{1}=-\epsilon_{2}$,
\begin{eqnarray}
I_{i}\approx\int d\omega\frac{8\lambda_{0}^{4}\Gamma_{0}^{2}
\left[f_{\omega}^{-}-f_{\omega}^{+}\right]\left[(\epsilon_{M}^{2}
+\omega^{2})(\epsilon_{0}^{2}-\omega^{2})^{2}+8\lambda_{0}^{2}
\omega^{2}(2\lambda_{0}^{2}+\epsilon_{0}^{2}-\omega^{2})
+8\epsilon_{0}^{2}\epsilon_{M}^{2}\omega^{2}\right]}{A_\omega^{2}+B_\omega^{2}},
\end{eqnarray}
\begin{eqnarray}
S_{12}\approx\int d\omega\frac{16\epsilon_{M}^{2}
\lambda_{0}^{4}\Gamma_{0}^{2}\left[(\epsilon_{0}^{2}
-\omega^{2})^{2}(f_{\omega}^{-}-f_{\omega}^{+})^{2}
+8\epsilon_{0}^{2}\omega^{2}(f_{\omega}^{-}+f_{\omega}^{+}
-2f_{\omega}^{-}f_{\omega}^{+})\right]}{\left[A_\omega+B_\omega^{2}/A_\omega\right]^{2}},
\end{eqnarray}
from which, we can obtain Eqs. (\ref{S12-asym-largeV}) and
(\ref{S12-asym-smallV}) for large and small lead voltages,
respectively.

\section{The Model with MBS replaced by a normal superconductor}\label{sec:non-Majorana}

For comparison, we also consider a non-Majorana setup. It consists
of a superconductor coupled to two quantum dots, and each dot is
connected to a normal metallic electrode. The Hamiltonian of the
double dots with on-site Coulomb repulsion is described by
\begin{eqnarray}\label{Hddot}
H_{DD}=\sum_{i\sigma}\epsilon_{i\sigma}d_{i\sigma}^\dagger d_{i\sigma}
+\sum_{i}U_in_{i\uparrow}n_{i\downarrow},
\end{eqnarray}
where where $i(=1, 2)$ represents the dot index and the on-site
Coulomb interactions are measured by $U_i$.  The Hamiltonian of the
superconducting lead reads
\begin{eqnarray}\label{Hsc}
H_{SC}=\sum_{k\sigma}\epsilon_{sk\sigma}c_{sk\sigma}^\dagger c_{sk\sigma}
+\sum_{k}\Delta (c_{sk\uparrow}^\dagger c_{s-k\downarrow}+\mathrm{H.c.})
\end{eqnarray}
with the lead-electron operators $c_{sk\sigma}^\dagger$ and
$c_{sk\sigma}$. The superconductivity in the lead is described by
the pairing order parameter $\Delta$. The electrochemical potential
of the superconductor is taken as the reference for energies and set
to zero, i.e., $\mu_S=0$. The Hamiltonian describing the tunneling
between the superconducting lead and the double dot is given by
\begin{eqnarray}\label{HtunnS}
H_{TS}=\sum_{ik\sigma}[t_{si}c_{sk\sigma}^\dagger d_{i\sigma}+\mathrm{H.c.}].
\end{eqnarray}
Assuming that the normal-state density of states in the
superconducting lead $\rho_s$ is constant in the energy window
relevant for transport, we define the coupling strength
$\Gamma_{si}=2\pi\rho_s|t_{si}|^2$ for the tunneling between the
superconductor and dot $i$. In the case that $\Delta$ is much
stronger than other system parameters, the quasiparticles in the
superconductor are inaccessible, and one can trace out of the
degrees of freedom of the superconducting lead without inducing any
dissipative dynamics. By performing a real-time perturbative
expansion and adding up all contributions in $\Gamma_{si}$, we
obtain the following effective Hamiltonian describing the dynamics
of the double dots\cite{Eldridge10PRB,Governale08PRB}
\begin{eqnarray}
H_{\mathrm{eff}}&=&H_{DD}-\sum_i\frac{\Gamma_{si}}{2}(d_{i\uparrow}^\dagger
d_{i\downarrow}^\dagger+\mathrm{H.c.})
+\frac{\sqrt{\Gamma_{s1}\Gamma_{s2}}}{2}
(d_{1\uparrow}^\dagger d_{2\downarrow}^\dagger-d_{2\uparrow}^\dagger
d_{1\downarrow}^\dagger+\mathrm{H.c.}),
\label{Heff}
\end{eqnarray}
where two new contributions are added to the double-dot Hamiltonian.
The first one describes the local Andreev reflection for each dot.
The second term describes the formation of nonlocal superconducting
correlations between the two dots induced by the splitting of Cooper
pairs into the two dots. In realistic devices, $U_{i}$ is of the
order of 1meV and is much larger than the other energy scales
relevant for transport. We calculate the cross correlation for this
setup in the limit $U_{i}\rightarrow\infty$ (so that each dot is
singly occupied) and the results are shown in Fig.
\ref{fig:E0-noMBS}.

\end{widetext}

\end{document}